\documentstyle[prl,aps]{revtex}
\newcommand{\be}{\begin{equation}}
\newcommand{\ee}{\end{equation}}
\newcommand{\ba}{\begin{eqnarray}}
\newcommand{\ea}{\end{eqnarray}}
\newcommand{\vq}{{\bf q}}
\newcommand{\vk}{{\bf k}}
\begin{document}
\title{Resonant Raman scattering of quantum wire in strong
magnetic field}
\author{Hyun C. Lee}
\address{Asia Pacific Center for Theoretical Physics, 130-012, 
Seoul, Korea}
\maketitle
\draft
\begin{abstract}

The resonant Raman scattering of a quantum wire in a strong magnetic field is studied, 
focused on the effect of long range Coulomb interaction and the spin-charge separation.
The energy-momentum dispersions of charge and spin excitation obtained 
from Raman cross-section show 
the characteristc cross-over behaviour induced by inter-edge Coulomb interaction.
The ``SPE" peak near resonance in {\em polarized} spectra 
 becomes broad due to the 
momentum dependence of charge velocity. The broad peak in the {\em depolarized }spectra 
is shown to originate
from the disparity between charge and spin excitation velocity. 
\end{abstract}
\thispagestyle{empty}

The edges of quantum Hall (QH) bars or the quantum wires (QW) in strong magnetic field 
\cite{wen,moon94} provide 
clean experimental verifications of many interesting theoretical predictions of 
Luttinger liquids \cite{mill,chang}. Recent studies have shown that 
the long range Coulomb interaction (LRCI) brings new effects into these systems
\cite{FB,OF,moon96,schulz,LY}. Basically LRCI causes a cross-over from Fermi/Luttinger liquid
phase to a phase where {\it inter-edge} Coulomb interactions dominates. An example of the 
cross-over in terms of the charge density wave correlation function is \cite{FB,schulz}
\begin{eqnarray}
\label{cdw}
C(x)\sim \cases{   (x/\ell)^{-2/\nu}, \;\;x<W \cr
\cos(2k_Fx) \exp(-\frac{2}{\nu}\sqrt{\alpha\ln(x^2 \beta)}), \;x>W, \cr} 
\end{eqnarray}
where $\ell$, $W$, and $k_F$ are the magnetic length, the width of the Hall
bar, and the Fermi wavevector.
As shown in Eq.({\ref{cdw}), the cross-over length scale is given by the width of the Hall bar.
The physical origin of the cross-over scale is
 the spatial separation of left and right moving electrons in 
QW in strong magnetic field, and it makes 
QW in strong magnetic field distinct from  ordinary QW.
Below the cross-over momentum scale, the {\em inter}-edge Coulomb interaction is effective, and the 
electron liquids become {\em non-chiral}. Above the scale the 
the {\em inter}-edge Coulomb interaction is negligible, and the electron liquids become 
{\em chiral} (left and right moving modes are decoupled).
The physical effects of the cross-over scale were first studied
in the edge states of Quantum Hall effects, and 
the tunneling conductances and the various density-density correlation functions
were found to be
drastically modified at low temperature and energy\cite{moon96,LY,nagaosa}.
The spin charge separation is  one of the  conspicuous features of 
one dimensional electronic system.  
The singular effects of LRCI do not influence the properties of spin excitations 
thanks to spin-charge separation\cite{note1}.

We have investigated the properties of the QW in strong magnetic field focusing on 
the effect of LRCI and the spin-charge separation in the light of Raman scattering. 
The energy-momentum dispersion relation of {\it charge} density excitations exhibits 
the singularity coming  from LRCI and the characteristic cross-over behaviour. 
The dispersion relation of {\it spin} density excitations shows {\it much weaker}
 cross-over behavior mainly
due to the lack of singular LRCI effect. 
A very unique feature of Raman scattering of quantum wire is that by adjusting the polarization 
and the resonance condition, the charge and spin density excitations can be probed selectively
\cite{sass98}.
The so-called ``SPE" peak which appears near resonance \cite{sass98} (see below)
is also found to be
strongly influenced by the LRCI and the spin-charge separation effect. 
The charge contribution to the ``SPE' peak in the {\em polarized } configuration is 
broadened by the momentum dependence of charge velocity, while the spin 
contribution to the ``SPE' peak in the {\em polarized } configuration remains sharp.
The  ``SPE' peak in the {\em depolarized } configuration becomes very broad, which is 
due to the {\it disparity } between charge and spin velocity. Basically, it is of the same origin
as that of the incoherent nature of electron Green functions in Luttinger liquid.
The  magnitude of Raman scattering cross section is enhanced or suppressed depending 
on the polarization configuration by a factor determining the cross-over. 
For explicit results, see Eq.(\ref{lowest},\ref{pol.col},\ref{spin},\ref{incoherent}).

The Raman spectroscopy is a very powerful probe in investigating the physical 
properties of one dimensional electronic systems. In particular, 
the dispersion relations of collective and single-particle excitations can be 
measured \cite{review}. 
Away from the resonance condition,  the charge density excitation has been 
identified in the {\em polarized } spectra in which the polarizations of 
incident photon and scattered photons are parallel, while the spin density
excitation has been identified in the {\em depolarized} spectra where 
the polariztions of incident and scattered photons are perpendicular.
Near the resonance, the additional structures("SPE" peak \cite{sass98})  
have been observed in both spectra,
and there have been several studies on the physical origin of the structures 
in the resonant Raman spectra
\cite{sass98,exp1,exp2,exp3,exp4,sarma,hwang}.
Especially, Sassetti and Kramer \cite{sass98} attributed the origin of 
the additional peak in the {\em polarized} Raman spectra  near the resonance 
to the collective 1D {\em intra-band  spin }excitations. The additional peak was shown to
 originate from the higher order term in the Raman cross section
related to the resonance condition. 
For the {\em depolarized } spectra, the broad peak feature of the 
scattered intensity as a function of frequency shift was predicted, 
which is due to the {\em simultaneous and independent} propagation
 of charge and spin excitaion.

In the standard theory of Raman spectroscopy, 
the differential Raman cross section is given by \cite{sass98,blum}
\ba
\label{section}
\frac{d^2 \sigma}{d \Omega d \omega}&=&\left( \frac{e^2}{m c^2} \right)^2\,
\frac{\omega_f}{\omega_i}\,\frac{n_\omega+1}{\pi}\,{\rm Im} \chi(\vq,\omega),\;\;
\chi(\vq,t)= i \theta(t)\,\langle [ N^\dag(\vq,t),N(\vq,0) ] \rangle,\nonumber \\
N(\vq)&=&\sum_{s,\vk}\,  \frac{\gamma_s}{D(\vk)}\,c^\dag_s(\vk+\vq)\,c_s(\vk),\;\;
n_\omega=\frac{1}{e^{\beta \omega}-1}, \;\;
\vq=\vk_i-\vk_f,\;\;\omega=\omega_i-\omega_f,
\ea
where and $c_s(\vk)$ is an electron
operator with momentum $\vk$ and spin $s$ and 
$\vk_{i(f)}, \;\; \omega_{i (f)},\;\;{\bf e}_{i (f)}$ 
are the momentum, energy, and polarization vector of the incident (scattered)
photons, respectively.
The coefficents of the scattering operators $N(\vq)$ are given by
\be
\label{coeff}
\gamma_s=\gamma_0\,\Big( {\bf e}_i \cdot {\bf e}_f + i s\,|{\bf e}_i \times {\bf e}_f
| \Big),\;\;
D(\vk)= E_c(\vk+\vq)-E_v(\vk+\vq-\vk_i)-\hbar \omega_i.
\ee
The Eq.(\ref{coeff}) assumes only one conduction and valence band. 
$E_c$ and $E_v$ are the conduction and valence band energies in the 
effective mass approximation. $\gamma_0$ is  the transition matrix element
between valence and conduction band, which is assumed to be constant.
Even in the presence of the external magnetic field the Eq.(\ref{section},\ref{coeff}) 
are valid.  In particular, the spin dependence remains intact because the Hamiltonian
of QW in a strong magnetic field conserves the z-component of the spin 
and because the denominator $D(\vk)$ depends only on the 
difference in energy between conduction and valence band. Therefore, if we neglect 
the band dependence of g-factor, the spin dependence remains intact.
The matrix element $\gamma_0$ depends on the orbital part of the single particle 
Hamiltonian, and it may acquire  momentum dependence, but we neglect 
them.


We adopt the following model for a quantum wire 
in strong magnetic fields  
(or equivalently a narrow Hall bar \cite{LY,YG}).
When the transverse motion is confined
by a parabolic potential the single-electron energy levels are given by 
$E_{n,k}=(n+1/2)\Omega_0+\frac{\hbar^2 k^2}{2m} $,
where the enchanced longitudinal mass is $m=m^* (\Omega_0/\omega_0)^2$. 
The subband energy spacing is  
$\Omega_0=\sqrt{\omega_c^2+\omega_0^2}$, 
where $\omega_c=eB/m^*c$ is the cyclotron energy
and $\omega_0$ is the frequency of the harmonic potential.
The electronic wavefunction at
the Fermi wavevectors $k_{F,r,s}$ of the lowest magnetic subband 
is given by 
\begin{equation}
\phi_{r,s}(x,y)=\frac{ e^{i k_{F,r,s} x}}{\pi^{1/4}\,\tilde{\ell}^{1/2}}\,
\exp\Big(-\frac{(y-R^2 k_{F,r,s})^2}{2\tilde{\ell}^2} \Big),
\end{equation}
where  
$\,\tilde{\ell}=(\hbar/m \Omega_0)^{1/2} \,$,
$R^2=\hbar \omega_c /( m \Omega_0^2) \,$, 
$r=R (L)$ for the right (left) branch,  
and $y$ is the transverse coordinate \cite{CP}. 
Because of the large 
longitudinal mass at fields where
$\omega_c> \omega_0$ all the electrons can be accommodated 
in the lowest magnetic subband. 
Each branch $r$ consists of 
spin-up and -down edges ($s=\uparrow,\downarrow$).
The intra branch LRCI are given by ($\gamma$ is Euler constant.)\cite{moon96,zulicke96}
\ba
& &V^{{\rm intra}}_{q,\parallel}=
-\frac{ 2 e^2}{\epsilon}\,\ln \frac{\gamma q a_{\parallel}}{2},\;\;
V^{{\rm intra}}_{q,\perp}=
-\frac{ 2 e^2}{\epsilon}\,\ln \frac{\gamma q a_{\perp}}{2},
\ea
and the inter branch LRCI are given by
\ba
& &V^{{\rm inter}}_{q,\parallel,s}=
\frac{ 2 e^2}{\epsilon}\,K_0(q W_{\parallel,s}),\;\;
V^{{\rm inter}}_{q,\perp}=
\frac{ 2 e^2}{\epsilon}\,K_0(q W_{\perp}).
\ea
Here $W_{\parallel,s}=|k_{F,R,s}-k_{F,L,s}|R^2$ and
$W_{\perp}=|k_{F,R,s}-k_{F,L,-s}|R^2$, which are order of the width of QW.
In the study of Raman scattering,
the small differences among $W_{\parallel,s},\,W_{\perp}$ are inessential, though 
they play a very important role in the inter-edge spin correlation functions and 
tunneling
conductances \cite{LY,Hyun}. 
$K_0(x)$ is 
the modified Bessel function, which behaves like 
$-\ln x$ in the limit of small $x$ and $e^{-x}/\sqrt{x}$ for large $x$. Therefore,
the inter-edge LRCI is negligible 
for the momentum  $q > \max(1/W_{\parallel,s}, 1/W_{\perp})$. The associated 
cross-over energy scale is given by $\omega_{{\rm cr}}=\frac{v_0}{W},\;\;
v_0=\frac{2 e^2 }{\hbar \pi \epsilon}$, 
where $W (\sim W_{\parallel,s} \sim W_{\perp}) $ is the width of QW.

It is convenient to use charge and spin 
density operators 
$\, \rho_{r,q}=\frac{\rho_{r,q,\uparrow}+
\rho_{r,q,\downarrow}}{\sqrt{2}} $ and
$\sigma_{r,q}=\frac{\rho_{r,q,\uparrow}-\rho_{r,q,\downarrow}}{\sqrt{2}}$.
The bosonized Hamiltonian can be written as a sum of the charge, spin, 
and mixed terms
$ H=H_\rho+H_\sigma+\delta H $,
where
\begin{equation}
H_\rho=\frac{\pi v_F}{L}\,\sum_{r,q \neq 0}\,
: \rho_{r,q}\,\rho_{r,-q}:+ 
\frac{1}{L}\,\sum_{r,q}\,V^{{\rm intra}}_{\rho,q}\,:\rho_{r,q}\,\rho_{r,
-q}: 
+ \frac{2}{L}\,\sum_q  V^{{\rm inter}}_{\rho,q}\, \rho_{R,q}\,\rho_{L,-q},
\end{equation}
\begin{equation}
H_\sigma=\frac{\pi v_F}{L}\,\sum_{r,q \neq 0}\,
:\sigma_{r,q}\,\sigma_{r,-q}:+ 
\frac{1}{L}\,\sum_{r,q}\,V^{{\rm intra}}_{\sigma,q}\,
:\sigma_{r,q}\,\sigma_{r,-q}: 
+\frac{2}{L}\,\sum_qV^{{\rm inter}}_{\sigma,q}\,\sigma_{R,q}\,\sigma_{L,-q},
\end{equation}
\begin{equation}
\label{mixing}
\delta H=2 \delta v_F\,\frac{\pi}{L}\,\sum_{r,q}\,\sigma_{r,q}\,\rho_{r,-q}
+\frac{1}{L}\,\sum_q\,\delta V^{{\rm inter}}_q\,\Big[\sigma_{R,q} \rho_{L,-q}
+\sigma_{L,q}\, \rho_{R,-q} \Big],
\end{equation}
\begin{eqnarray}
& &V^{{\rm intra}}_{\rho (\sigma),q}=
\frac{V^{{\rm intra}}_{q,\parallel}\pm
V^{{\rm intra}}_{q,\perp}}{2},\;\;
V^{{\rm inter}}_{q,\parallel}
=\frac{V^{{\rm inter}}_{q,\parallel,\uparrow}
+V^{{\rm inter}}_{q,\parallel,\downarrow}}{2}, \;\;\delta V^{{\rm inter}}_q=
\frac{V^{{\rm inter}}_{q,\parallel,\uparrow}
-V^{{\rm inter}}_{q,\parallel,\downarrow}}{2},
\nonumber \\
& &V^{{\rm inter}}_{\rho (\sigma),q}=
\frac{V^{{\rm inter}}_{q,\parallel}\pm
V^{{\rm inter}}_{q,\perp}}{2},\;\;
v_F=\frac{v_{F\uparrow}+v_{F \downarrow}}{2},\;\;
\delta v_F=\frac{v_{F\uparrow}-v_{F \downarrow}}{2}.
\end{eqnarray}
Note that in the effective Hamiltonian, 
the charge and spin degrees of freedom are 
separated  
except in $\delta H$ \cite{LY}.
It is convenient to introduce 
\begin{eqnarray}
v_{\rho \pm}&=&v_F+\frac{V^{{\rm intra}}_{\rho,q}}{\pi}\pm
\frac{V^{{\rm inter}}_{\rho,q}}{\pi},\,\, \xi_\rho(q)\equiv
\left(\frac{v_{\rho +}(q)}{v_{\rho-}(q)}\right)^{1/2},\;\;
v_\rho(q)\equiv\sqrt{v_{\rho +}(q) v_{\rho -}(q)}
 \nonumber \\
v_{\sigma \pm}&=&v_F+\frac{V^{{\rm intra}}_{\sigma,q}}{\pi}
\pm\frac{V^{{\rm inter}}_{\sigma,q}}{\pi},\,\,\xi_\sigma(q)\equiv
\left(\frac{v_{\sigma +}(q)}{v_{\sigma -}(q)}\right)^{1/2}
,\;\;
v_\sigma(q)\equiv\sqrt{v_{\sigma +}(q) v_{\sigma -}(q)}.
\end{eqnarray}
In the high momentum regime $q> 1/W$  where the inter-edge LRCI can be neglected,
the velocities are given by
\ba
\label{vel1}
& &v_{\rho +}=v_{\rho -}=v_\rho=v_F+v_0\, 
\ln \frac{2}{\gamma |q| \sqrt{a_\parallel a_\perp}},\;\;\xi_\rho(q)=1 \nonumber \\
& &v_{\sigma +}=v_{\sigma  -}=v_\sigma=v_F+v_0\, \ln \Big(
\frac{a_\perp}{a_\parallel}\Big)^{1/2}\sim v_F,\;\;\xi_\sigma(q)=1.
\ea
In the low momentum regime $q < 1/W$ where  the inter-edge LRCI  is effective,
the velocities are given by
 ( $W_\parallel=\big( W_{\parallel \uparrow }
W_{\parallel \downarrow} \big)^{1/2}$),
\ba
\label{vel2}
& &v_{\rho +}=2 v_0\,\ln \frac{1}{|q|d},\;\;\,
v_{\rho -}= v_F+ v_0\, \ln \Big( \frac{W_\parallel W_\perp}{a_\parallel a_\perp} \Big)^{1/2},\;\;
v_\rho \sim v_0 \sqrt{\ln \frac{1}{|q|d}},\;\;
 \xi_\rho(q) \propto  \sqrt{\ln \frac{1}{|q|d}}
\nonumber \\
& &v_{\sigma +}= v_F+ v_0\, \ln \Big( \frac{a_\perp W_\perp}{a_\parallel W_\parallel} \Big)^{1/2},
\;\;
v_{\sigma -}= v_F+ v_0\, \ln \Big( \frac{W_\parallel a_\perp}{a_\parallel   W_\perp} \Big)^{1/2},
\;\;\xi_\sigma \sim 1,\;\;v_\sigma \sim v_F,
\ea
where $ d = \frac{\gamma}{2}\,e^{-\frac{v_F}{2 v_0}}
(W_\parallel W_\perp a_\parallel a_\perp)^{1/4}$. 
Note the cross-over behaviour of the charge velocity $v_\rho(q)$ and the inter-edge Coulomb factor
$\xi_\rho(q)$.  The spin velocity $v_\sigma$ and $\xi_\sigma$ is very weakly 
 renormalized by 
the Coulomb interactions since the differences among the length scales are negligible.

Next we turn to the computation of the correlation function
$ \chi(q,t)$. 
Away from the resonance, namely, $\hbar v_F q << |E_G-\hbar \omega_i|$ and 
$\hbar \omega << |E_G-\hbar \omega_i|$, the momentum dependence of $D(\vk)$ can 
be neglected. Here, $E_G$ is the distance between the conduction and 
valence band at the Fermi wave vector in the effective mass approximation.
Then the scattering operator $N(\vq)$ can be simplified to \cite{sass98}
\ba
\label{classical}
N(\vq)&=&\frac{\gamma_0}{E_G-\hbar \omega_i}\,
\Big[{\bf e}_i \cdot {\bf e}_f \,\rho(\vq) +
     i|{\bf e}_i \times {\bf e}_f | \sigma(\vq) \Big],\nonumber \\
\rho(\vq)&=&\rho_\uparrow(\vq)+ \rho_\downarrow(\vq),\;\;
\sigma(\vq)=\rho_\uparrow(\vq)- \rho_\downarrow(\vq).
\ea
Depending on the mutual polarizations (parallel, or perpendicular) of the incident
and scattered photons, only the charge or spin density operator contributes 
to the Raman cross section, respectively.
This is  the {\em classical } selection rule of Raman spectra of quantum wires 
and dots, which is valid in the lowest order of 
$\frac{\hbar v_F q }{|E_G-\hbar \omega_i|},\;\;
\frac{\hbar \omega }{|E_G-\hbar \omega_i|}$\cite{note2}.
Close to the
resonance such that $\hbar \omega_i \approx E_G+ \hbar v_F q$ the momentum
dependence of $D(k)$ should be taken into account. Following Sassetti and Kramer
\cite{sass98}
we expand the  $\frac{1}{D(k)}$ to include the momentum dependence (assuming
back scattering $k_i=q/2$ and the linerization approximation), 
we get the two leading  contributions to the scattering operator $N(q)$.  
\ba
\Delta N_1^{(r)}(q)&=&-\frac{ \hbar \eta}{(E_G-\hbar \omega_i)^2}\,\sum_{s,k}\,
\gamma_s\,v_{F}\,(r k - k_{F}) 
\, c_s^{(r)\dag }(k+q)\,c_s^{(r)}(k),\quad r=R,L=\pm1, \\
\Delta N_2^{(r)}(q)&=&-\frac{ \xi v_F q}{(E_G-\hbar \omega_i)^2}\,
\sum_s\,\gamma_s \,c_s^{(r)\dag }(k+q)\,c_s^{(r)}(k). 
\ea
For the explicit computations of the correlation function 
$ \chi(q,t)$, the above operators 
should be bosonized.
The operator $\Delta N_2^{(r)}(q)$ can be  bosonized immediately , and 
$\Delta N_1^{(r)}(q)$ can be bosonized by using the commutation relation
$\big[ \rho_{r s}(q),\rho_{r s}(q^\prime) \big]=
-\frac{r q L}{2\pi}\,\delta_{q+q^\prime,0}$.
In terms of the charge and spin density operators, 
the bosonized  $\Delta N_1(q)=\sum_r\,\Delta N_1^{(r)}$ reads\cite{sass98}
\be
\Delta_1 N(q)=\sum_{r, k}\,\frac{(-1)\eta \,v_F\,
\gamma_0}{(E_G-\hbar \omega_i)^2}\,\frac{\pi}{2 L}\,\Big[
 {\bf e}_i \cdot {\bf e}_f
 \,\big(:\rho^{(r)}_{k}\,\rho^{(r)}_{q-k}: +(\rho \to \sigma)\big)+2i
 |{\bf e}_i \times {\bf e}_f|\,:\rho^{(r)}_k\,\sigma^{(r)}_{q-k}:\Big],
\ee
where $: :$ denotes the normal ordering with respect to the filled Fermi sea.

First, consider the polarized configuration, ${\bf e}_i \cdot {\bf e}_f=1$.
In the lowest order of the $\frac{\hbar v_F q}{E_G-\hbar \omega_i}$, only the charge 
density operator contributes to the correlation function $\chi(q,\omega)$ (See Eq.(\ref{classical})).  
The correlation function can be computed straightforwardly.
\be
\label{lowest}
{\rm Im} \chi_\rho(q,\omega)=\frac{L\,\gamma_0^2\,q}{(E_G-\hbar \omega_i)^2}\,
\xi_\rho^{-1}(q)\,\delta(\omega- v_\rho(q) q).
\ee 
The energy-momentum dispersion is given by $\omega = v_\rho \, q$, and it should exhibit
the cross-over 
\begin{eqnarray}
v_\rho(q)\sim \cases{  v_0 \, \ln \frac{const.}{|q|}, \;\;q > W, \cr
          v_0 \,\sqrt{\ln \frac{1}{|q| d}},\;\; q < W, \cr}. 
\end{eqnarray}
The dispersion in the Coulomb dominated regime is  of the plasmon type 
\cite{sass98,exp1,gold,li}.One
 notable feature is the {\em suppression} of the scattering intensity at low momentum
$q < 1/W $ due to the inter-edge Coulomb factor $\xi_\rho^{-1}(q)$.
The typical width of the quantum wire is [ $W\sim750$\AA ]\cite{exp1}, which is 
 $1 \times 10^{5}\,{\rm cm}^{-1}$ in the wave number units, and it means  
the inter-edge LRCI is operative for $q < 1 \times 10^{5}\,{\rm cm}^{-1}$.
The typical  momentum transfer in Raman scattering process of QW is also 
of the order of $1 \times 10^{5}\,{\rm cm}^{-1}$ \cite{exp1}. 
Therefore, the cross-over into the 
Coulomb dominated regime could be observed  in the Raman scattering experiment.

Near the resonance, the contributions from
the correlation functions  of the additional scattering operators 
$\Delta N_i(q) $ have to be included. The contributions from the charge part of $\Delta N_1$ is
\ba
\label{charge}
{\rm Im}\chi^{(1)}_\rho(q,\omega)&=&\frac{L (\eta \hbar v_F \gamma_0)^2}{ (E_G-\hbar \omega_i)^4}
\,\sum_{0<p<q}\,p (q-p)\,(\xi_\rho(p)+\frac{1}{\xi_\rho(p)})\,\nonumber \\
&\times&\delta (\omega-q v_\rho(q-p)-p (v_\rho(p)-v_\rho(q-p))).
\ea
If the velocity of charge excitation $v_\rho$ were momentum independent like in Luttinger 
model, the delta function part would simplify into 
$\delta(\omega - v_\rho q)$ which is independent of $p$,and the  sharp delta funtion 
type  peak  would appear as in Eq.(\ref{lowest}).
However, due to the momentum dependence, the sharp peak is broadened.
If $q < 1/W$,
\be
\label{pol.col}
{\rm Im}\chi^{(1)}_\rho(q,\omega) \sim
p_\rho^*(q-p_\rho^*)\,\max(\xi_\rho(p_\rho^*),\xi_\rho(q-p_\rho^*)),\;\; {\rm for}\;\;
\bar{v}_0 q \sqrt{\ln \frac{1}{qd}}< \omega < \bar{v}_0 q \sqrt{\ln \frac{2}{qd}},
\ee
where $\bar{v}_0 \sim \frac{ 2 e^2}{ \hbar \pi \epsilon}\,\sqrt{\ln \frac{W}{a}},\;
p_\rho^*=\frac{\omega/\bar{v}_0}{\sqrt{\ln \frac{\bar{v}_0}{\omega d}}}$.
Note that the scattering intensity is {\em enhanced} by the inter-edge LRCI 
factor $\max(\xi_\rho(p_\rho^*),\xi_\rho(q-p_\rho^*))$. 
In the polarized configuration there also exists the contribution from the spin 
sector. In constrast to the charge excitation velocity, 
the spin excitation velocity is independent of
momentum once the momentum is lower than $1/W$.
(See Eq.(\ref{vel1},\ref{vel2})) 
 As a consequence, the broadening 
from the momentum dependence of velocity is absent .
If $ q < 1/W$,
\be
\label{spin}
{\rm Im}\chi^{(1)}_\sigma(q,\omega)=\frac{L (\eta \hbar v_F \gamma_0)^2}{ (E_G-\hbar \omega_i)^4}
\,q^3\,\delta(\omega-q v_\sigma)\,\big(\xi_\sigma+\xi_\sigma^{-1} \big).
\ee
which is qualitatively different from the charge  contribution 
(\ref{pol.col}).
$\xi_\sigma+\xi_\sigma^{-1} $ is a contribution from the inter-edge interaction, and 
it is constant.
Accordingly, even for the QW in a strong mangetic field in the Coulomb dominated regime, 
the "SPE" peak\cite{sass98} near resonance exists, 
which is solely due to the collective intra-band
{\em spin} excitations. In passing we note that
 the sharp peak would appear {\em both} in charge and spin channel in Luttinger liquid.
The contributions from the operator $\Delta N_2(q)$ is
\be
{\rm Im}\chi^{(2)}(q,\omega)=\frac{L \gamma^2 (\xi v_F q)^2\,q}{(E_G-\hbar \omega_i)^4}\,
\xi_\rho(q)\,\delta(\omega -v_\rho q).
\ee
It is similar to ${\rm Im}\chi_\rho(q,\omega)$, but the Coulomb interaction 
{\em enhances} the Raman scattering cross section.

Second, consider  the depolarzied configuration 
$|{\bf e}_i \times {\bf e}_f|=1$.
In the lowest order of 
$\frac{\hbar v_F q}{E_G-\hbar \omega_i}$, only the spin operator contributes to the correlation
function.
\be
{\rm Im} \chi^{(0)}(q,\omega)=\frac{L\,\gamma_0^2\,|q|}{(E_G-\hbar \omega_i)^2}\,
\xi_\sigma^{-1}(q)\,\delta(\omega-v_\sigma |q|).
\ee
The spin excitation velocity is constant except near the cross over region 
$q \sim 1/W$. In most other region, the dispersion is Luttinger like 
$\omega = v_\sigma q$. There is a also inter-edge LRCI factor $\xi_\sigma^{-1}(q)$
but its effect is almost negligible in the present context \cite{LY} and it is constant.
The additional contriubution from $\Delta N_1(q)$ near the resonance gives 
the broad peak in a certain
range of frequency {\em both } in the Luttinger \cite{sass98} and in the Coulomb dominated cases 
as ours.
The broad peak originates  from the {\em disparity} between  charge and spin velocity and from
the {\em simultaneous and independent}
 propogation of charge and spin excitations. 
 The origin of the broad peak is  also related to
the incoherent nature of the electron Green function in Luttinger model,but it has nothing to 
do with that of ${\rm Im} \chi^{(1)}_\rho(q,\omega)$ in the polarized configuration 
[see Eq.(\ref{pol.col})].
In the Couloumb dominated regime,
the frequency range of broad peak ( $q< 1/W$) is
$q v_\sigma \le \omega \le q \bar{v}_0 \, \sqrt{ \ln \frac{1}{qd}}$.
Note that the lower bound is given by spin excitation energy scale, while the upper bound 
is given by LRCI scale, and this clearly demonstrates  the origin of broad peak explained above.
For $\omega \sim q v_\sigma$, an explicit result of the correlation function is given
by
\be
\label{incoherent}
{\rm Im} \chi^{(1)}(q,\omega) \sim 
\frac{L (\eta \hbar v_F \gamma_0)^2}{ (E_G-\hbar \omega_i)^4}\,
p_0\,(q-p_0) \,
\max(\xi_\rho(p_0),\xi_\rho(q-p_0))+\max(\xi_\sigma(p_0),\xi_\sigma(q-p_0)),
\ee
where $p_0=\frac{(\omega-q v_\sigma)/\bar{v}_0}{
\sqrt{\ln \frac{\bar{v}_0}{(\omega-q v_\sigma) d}}}$.
Note the Coulomb enhancement factor is the sum of that of charge and spin contribution.

In summary, we have studied the effect of LRCI on the resonant Raman scattering 
of QW in a strong mangetic field. Due to the spatial 
separation of left and right moving electrons in strong magnetic fields 
a new Coulomb momentum ( and energy) scale
emerges, and the Raman scattering clearly reveals the scale and the associated effects.
 Because the typical Coulomb 
cross-over momentum scale is the same order as the typical momentum transfer in Raman process,
our predictions can be verified 
experimentally.

The author has been supported by KOSEF.




\begin{references}
\bibitem{note1} More precisely, the spin sector is slightly modified by LRCI due to the 
shift of the Fermi vector induced by Zeeman splitting. But the modification is negligible 
in the study of Raman scattering. For details, see \cite{LY}.
\bibitem{note2} Strictly speaking, the classical selection rule does not hold
for QW in a magnetic field due to the charge-spin mixing term in the Hamiltonian
Eq.(\ref{mixing}). However, the contribution from the mixing term to the Raman 
cross section is the order of $(\delta v_F/v_F)^2$ which is negligible.
\end{references}
\end{document}